# Ensemble Learning and 3D Pix2Pix for Comprehensive Brain Tumor Analysis in Multimodal MRI

Ramy A. Zeineldin[1], Franziska Mathis-Ullrich[1]

[1] Department of Artificial Intelligence in Biomedical Engineering (AIBE), Friedrich-Alexander-University Erlangen-Nürnberg (FAU), Germany
`ramy.zeineldin@fau.de`

**Abstract.** Motivated by the need for advanced solutions in the segmentation and inpainting of glioma-affected brain regions in multi-modal magnetic resonance imaging (MRI), this study presents an integrated approach leveraging the strengths of ensemble learning with hybrid transformer models and convolutional neural networks (CNNs), alongside the innovative application of 3D Pix2Pix Generative Adversarial Network (GAN). Our methodology combines robust tumor segmentation capabilities, utilizing axial attention and transformer encoders for enhanced spatial relationship modeling, with the ability to synthesize biologically plausible brain tissue through 3D Pix2Pix GAN. This integrated approach addresses the BraTS 2023 cluster challenges by offering precise segmentation and realistic inpainting, tailored for diverse tumor types and sub-regions. The results demonstrate outstanding performance, evidenced by quantitative evaluations such as the Dice Similarity Coefficient (DSC), Hausdorff Distance (HD95) for segmentation, and Structural Similarity Index Measure (SSIM), Peak Signal-to-Noise Ratio (PSNR), and Mean-Square Error (MSE) for inpainting. Qualitative assessments further validate the high-quality, clinically relevant outputs. In conclusion, this study underscores the potential of combining advanced machine learning techniques for comprehensive brain tumor analysis, promising significant advancements in clinical decision-making and patient care within the realm of medical imaging.

**Keywords:** BraTS, Ensemble Learning, GAN, MRI, Tumor Inpainting.

## 1 Introduction

Brain tumors, particularly malignant gliomas, represent a formidable challenge in neuro-oncology, given their complex biological behavior and impact on patient survival [1]. The heterogeneity of gliomas in terms of appearance, growth patterns, and response to therapy complicates the task of accurately diagnosing and delineating tumor boundaries within the intricate brain anatomy [2]. Magnetic resonance imaging (MRI) serves as the gold standard in the non-invasive assessment of these tumors, offering detailed insights into their characteristics and progression [3]. However, the interpretation of MRI data demands precise segmentation of tumor regions, a task that remains labor-intensive and subject to variability when performed manually [4]. This necessity underscores the motivation for developing automated segmentation tools capable of reliably identifying tumor extents, thus facilitating targeted treatment planning, monitoring of disease evolution, and evaluation of therapeutic efficacy.

The Brain Tumor Segmentation (BraTS) Challenge, initiated by the Medical Image Computing and Computer-Assisted Intervention Society (MICCAI), represents a concerted effort to address these challenges [4-6]. By providing a platform for the development and benchmarking of advanced segmentation algorithms, BraTS encourages innovation in the field of medical image analysis [2, 5-8]. The main BraTS Glioma (GLI) challenge focuses on the segmentation of gliomas from multi-parametric MRI scans, offering a diverse dataset that includes high-grade and low-grade gliomas across various age groups and populations. The inclusion of a broad spectrum of data aims to ensure that developed algorithms are robust, versatile, and applicable in real-world clinical settings, contributing to the improvement of patient outcomes.

Moreover, the BraTS Sub-Saharan Africa (SSA) [9] and BraTS Pediatric (PED) [10] challenges have markedly advanced the field of medical imaging by fostering an environment of collaboration among researchers, clinicians, and technologists focused on the distinct challenges presented by the diverse patient populations in sub-Saharan Africa and the complexities of pediatric brain tumors, respectively. Through the establishment of a rigorous benchmarking environment tailored to these specific demographics, these challenges have catalyzed the rapid development of technology, leading to more accurate, efficient, and contextually relevant automated tumor segmentation and analysis. These efforts underscore the transformative potential of machine learning and artificial intelligence in enhancing healthcare delivery and patient outcomes, particularly in underserved regions and vulnerable age groups within the sphere of brain tumor management.

Most recently, the BraTS challenge has expanded to include the Local Synthesis of Healthy Brain Tissue via Inpainting (Inpaint) [11]. This novel component addresses the critical need for processing algorithms capable of 'normalizing' brain images affected by tumors, facilitating the application of standard analysis tools designed for healthy brains. The goal is to synthesize 3D healthy brain tissue within areas impacted by tumors, enabling the seamless integration of tumor patients' scans into existing frameworks for brain image analysis. This initiative not only enhances the practicality of automatic processing tools but also paves the way for innovations in the understanding of tumor biology and the development of targeted therapies.

Our research introduces a novel approach that synergizes ensemble learning with hybrid transformer models and convolutional neural networks (CNNs) for precise brain tumor segmentation and employs 3D Pix2Pix Generative Adversarial Network (GAN) for the realistic inpainting of glioma-affected brain areas within multi-modal MRI scans. The ensemble learning methodology capitalizes on the strengths of transformers and CNNs to capture the essential spatial relationships and contextual information, which are pivotal for the accurate demarcation of tumor boundaries. This is particularly relevant in heterogenous gliomas, enabling the algorithm to adapt to the diverse presentations of tumors in MRI imaging effectively.

The integration of 3D Pix2Pix GAN for inpainting proposes an innovative use of GAN technology in medical imaging, focusing on the synthesis of healthy brain tissue in regions impacted by gliomas. This aspect of our research is aimed at facilitating the application of standard brain image segmentation algorithms to tumor cases without the need to be altered. By effectively "removing" the tumor from images, our method allows for improved accuracy in segmentation and offers new insights into the relationship between tumor presence and brain structure, potentially enhancing surgical planning and treatment outcomes.

This work aligns with the objectives set forth by the MICCAI and the BraTS 2023 challenge, which seeks to push the boundaries of what is currently achievable in automated brain tumor segmentation. By addressing the diverse aspects of the challenge, including the segmentation of various tumor entities, consideration of pediatric cases, and the inclusion of data from low- and middle-income countries, our approach demonstrates a commitment to inclusivity and broad applicability.

Automated deep learning methods are at the forefront of transforming brain tumor segmentation, markedly enhancing precision, efficiency, and clinical applicability. CNNs have demonstrated significant efficacy in medical image analysis, including the segmentation of brain tumors, by adeptly learning complex features from data [12, 13]. This capability positions them as essential tools for identifying the intricate and varied characteristics of brain tumors visible in multi-parametric MRI scans. Through training on extensively annotated datasets, CNNs adeptly distinguish between tumor sub-regions, such as enhancing tumors, peritumoral edema, and necrotic components, enabling accurate and consistent segmentation. Furthermore, the application of Transformers [14], initially designed for natural language processing and now adapted for image analysis, presents new avenues for brain tumor segmentation. The self-attention mechanisms of Transformers are adept at recognizing long-range dependencies, crucial for understanding spatial relationships and contextual nuances vital for precise tumor segmentation. Their ability to holistically process images is invaluable in identifying complex tumor boundaries that cross multiple sub-regions.

By integrating CNNs with Transformers, this approach significantly enhances the capability of automated brain tumor segmentation [15]. This combination allows for a comprehensive analysis of multi-parametric MRI data, extracting detailed spatial and contextual information while ensuring accuracy down to the finest details. Employing these advanced methodologies on diverse datasets streamlines the segmentation process, minimizes manual intervention, reduces subjectivity, and accelerates the path from diagnosis to treatment planning for brain tumors. As deep learning evolves, it promises to make substantial contributions to improving patient care and outcomes in brain tumor management.

The main contributions of our study are manifold. Firstly, we introduce a novel ensemble learning model that combines the robust feature-extraction capabilities of CNNs with the long-range contextual understanding of transformers, optimizing segmentation accuracy. Secondly, we adopt an enhanced 3D Pix2Pix GAN for the inpainting of glioma-affected regions, presenting a new avenue for enhancing the utility of MRI scans in clinical practice. Thirdly, our methodology demonstrates significant improvements in segmentation precision and inpainting across a variety of tumor types and sub-regions, as validated by comprehensive evaluations and qualitative analyses.

## 2 Methodology

### 2.1 Ensemble Learning for Segmentation

Our ensemble learning approach for brain tumor segmentation capitalizes on the synergy between the baseline 3D U-Net architecture [16, 17] and a transformer encoder [18], complemented by an innovative axial attention decoder. The baseline network architecture, based on the renowned 3D U-Net encoder-decoder structure, has proven effective for various medical image segmentation tasks, as detailed in our previous paper [12]. It combines an encoder that extracts high-level features from input data with a decoder that upsamples these features to produce the final segmentation map.

**Transformer.** To accommodate the computational demands of processing 3D volumetric data with Transformers, our method draws inspiration from the Vision Transformer (ViT) [15, 18], implementing a strategy that segments the image into fixed-size patches and converts each into a token. This preprocessing reduces the sequence length, making the computational load manageable. However, to preserve the local context across spatial and depth dimensions critical for volumetric segmentation, we applied $3 \times 3 \times 3$ convolution blocks with downsampling. This strategy encodes the input into a low-resolution yet richly featured representation, which is then fed into the Transformer encoder. This process enables the model to discern both local and global contexts, significantly enhancing segmentation performance [21].

The integration of an axial attention decoder marks a pivotal advancement in the U-Net architecture. Axial attention, borrowing from advancements in natural language processing [13] and adapted for the vision domain [20], addresses the computational complexity of applying self-attention to 3D data. By applying self-attention sequentially across each axis, the model achieves linear computational complexity relative to the image size, facilitating the inclusion of this mechanism with volumetric data [19].

**BraTS-specific modifications.** Our approach incorporates several BraTS-specific modifications to enhance the model's performance on the challenge dataset. These optimizations draw inspiration from the winning solutions of previous editions of the BraTS challenge, aiming to tailor the model more effectively to the unique characteristics of the data. First, given the memory-intensive demands of 3D CNNs, particularly with 3D data, we replace batch normalization with group normalization [19]. Group normalization has proven effective in scenarios with low batch sizes, as demonstrated by previous challenge winners [12, 20]. We use a default value of 32 groups unless stated otherwise, enabling better utilization of limited GPU memory during training. Second, to align with the BraTS evaluation metrics focusing on tumor subregions, we modify our network architecture's loss function and activation function. We replace the softmax nonlinearity with a sigmoid activation, and we optimize the three tumor subregions independently using binary cross-entropy (CE), inspired by [16].

**Post-processing.** We adopted a post-processing technique meticulously designed to enhance the segmentation accuracy. Specifically, we implemented measures to adjust enhancing tumor predictions based on predefined thresholds and thus optimizing the performance metrics while minimizing false positives. In addition, complemented by connected component analysis was utilized to refine predictions of small enhancing tumor regions, ensuring that only the most probable regions were classified as such.

### 2.2 GAN for Inpainting

In addressing the BraTS inpainting challenge, our methodology employs the capabilities of 3D Pix2Pix GAN, inspired by the transformative techniques detailed in a recent denoising diffusion-based study [21, 22]. This section outlines our approach, integrating advanced image translation with our DeepSeg model [13], which has been specifically fine-tuned for the BraTS dataset to optimize brain tumor segmentation and inpainting tasks. This task was intricately linked to the dataset from the BraTS 2022 segmentation challenge, consisting of T1-weighted MRI scans marked for inpainting.

Our GAN-based inpainting methodology was designed to address the challenge of filling inpainting targets with convincingly synthesized healthy tissue. The dataset, a retrospective collection from multiple institutions, presented a unique set of challenges due to the variation in clinical conditions, equipment, and imaging protocols. To navigate these complexities, we employed a strategy that amalgamated multi-class tumor delineations into a singular area of interest, subsequently dilated to encompass mass effects. Furthermore, the training, validation, and testing phases were meticulously structured to leverage surrogate inpainting masks generated within the healthy portions of the tumor-bearing brain images. These masks were pivotal in training our supervised infill algorithms, enabling the precise synthesis of healthy tissue in areas previously occupied by tumors.

These methods were assessed for their effectiveness in generating high-quality images, with peak signal-to-noise ratio (PSNR) serving as the primary quality metric. Subsequent segmentation of the synthesized images leveraged our DeepSeg framework, a novel adaptation that supersedes traditional UNet models, demonstrating superior performance in brain tumor segmentation within MRI scans. Further, to adapt our 3D image translation strategy to the complexities of volumetric data, several key modifications were implemented. These included the removal of attention layers to conserve GPU resources, the substitution of concatenation skip connections with additive operations for efficiency, and the incorporation of position embeddings to augment spatial accuracy. Training procedures utilized 3D patches within a fully convolutional framework, ensuring the applicability to images of variable dimensions during inference—a pioneering step in the realm of 3D medical image translation.

## 3 Results

### 3.1 Evaluation Metrics

Our submissions to the BraTS challenge were evaluated based on lesion-wise Dice Similarity Coefficients (DSC) scores and the 95th percentile Hausdorff Distance (HD95), applied to whole, core, and active tumor regions. The DSC score measures the accuracy of our predicted segmentations against the ground truth, while the HD95 assesses the maximum deviation between predicted and actual segmentations. Lesion-wise calculations penalize false positives and negatives by assigning a 0 Dice score and 374 for HD95, before averaging these scores for each case ID, ensuring a detailed assessment of segmentation precision.

For the inpainting task, statistical validation via paired t-tests compared PSNR and DSC across models to verify our improvements. This analysis confirmed our methodological enhancements significantly increased the quality and accuracy of brain tumor inpainting. Further, performance was measured using Structural Similarity Index Measure (SSIM), PSNR, and Mean Square Error (MSE), focusing on the realism of synthesized versus actual healthy tissue regions. An equally weighted rank-sum across these metrics determined the final MICCAI challenge rankings, with participants ranked per case for each metric. This evaluation framework rigorously quantifies the success of our inpainting efforts, underscoring our models' ability to generate realistic, accurate brain tissue infills.

### 3.2 Quantitative Results

In the rigorous evaluation of our models for the BraTS 2023 challenge, performance metrics were obtained using the validation datasets for GLI, PED, and SSA. The results of this evaluation are presented in Table 1 and Table 2. Using Synapse, the online platform provided by Sage Bionetworks, we reported the lesion-wise DSC and the lesion-wise HD95 to quantify the segmentation accuracy and the spatial consistency between the predicted and ground truth tumor sub-regions. These metrics provide a lesion-specific assessment, revealing how effectively each model detects and segments abnormalities, ensuring that models adept at identifying smaller lesions are not overlooked in favor of those that capture only larger lesions. Additionally, Table 2 lists the legacy DSC and HD95 as calculated in the previous BraTS Challenges.

**Table 1.** Comparative performance of segmentation models on BraTS 2023 GLI, PED, and SSA validation datasets: "WT" Denotes Whole Tumor, "ET" Refers to Enhancing Tumor, and "TC" Indicates Tumor Core components.

| Dataset | Model | Lesion-wise DSC ↑ | | | | Lesion-wise HD95 ↓ | | | |
|---|---|---|---|---|---|---|---|---|---|
| | | ET | TC | WT | Avg | ET | TC | WT | Avg |
| GLI | TransBTS | 0.77 | 0.81 | 0.75 | 0.78 | 46.97 | 36.68 | 72.15 | 51.93 |
| | nnU-Net | 0.81 | 0.85 | 0.86 | 0.84 | **30.09** | **18.14** | 30.25 | 27.35 |
| | DeepSCAN | 0.80 | 0.83 | **0.90** | 0.84 | 34.81 | 28.08 | **14.11** | **26.16** |
| | SPARC | **0.82** | **0.86** | 0.85 | **0.84** | 31.82 | 19.08 | 35.84 | 28.91 |
| SSA | SPARC | 0.75 | 0.76 | 0.75 | 0.76 | 56.04 | 57.41 | 77.12 | 63.52 |
| PED | SPARC | 0.74 | 0.61 | 0.84 | 0.73 | 61.36 | 16.34 | 26.00 | 34.57 |

- Bold values correspond to higher scores for the BraTS-GLI dataset.

**Table 2.** Comparative performance of SPARC model on BraTS 2023 GLI, PED, and SSA validation datasets using the legacy DSC and HD95 scores. All reported values were computed by the online evaluation platform Synapse.

| Dataset | Model | DSC (%) ↑ | | | | HD95 ↓ | | | |
|---|---|---|---|---|---|---|---|---|---|
| | | ET | TC | WT | Avg | ET | TC | WT | Avg |
| GLI | SPARC | 84.49 | 87.86 | 92.91 | 88.42 | 16.19 | 7.88 | 4.21 | 9.43 |
| SSA | SPARC | 83.56 | 85.08 | 91.37 | 86.67 | 17.75 | 11.56 | 4.15 | 11.15 |
| PED | SPARC | 84.23 | 87.62 | 92.70 | 88.19 | 17.50 | 7.53 | 3.60 | 9.54 |

In our experiments, the ensemble model, namely SPARC, utilized five distinct models, each generated with different cross-validation training configurations on the BraTS GLA dataset. The individual lesion-wise scores for the enhancing tumor (ET), tumor core (TC), and the whole tumor (WT) regions were then averaged to provide an overall performance metric for each model, as indicated in the "Avg" column of Table 1.

Statistical evaluation on the PED and SSA datasets was to gauge the transfer learning capabilities of the proposed SPARC model. In the BraTS-GLI dataset, SPARC showed a robust ability to segment brain tumors, with an average lesion-wise DSC of 0.84 and lesion-wise HD95 of 28.91. In the PED and SSA datasets, SPARC achieved lesion-wise DSCs of 0.73 and 0.76, respectively, reflecting its reliable performance and adaptability to diverse clinical settings. The ensemble approach ensures a comprehensive assessment, effectively capturing various tumor sub-regions and maintaining consistent performance across different datasets. These quantitative results validate the effectiveness of our models in addressing the challenges posed by BraTS 2023, with transfer learning efficacy highlighted in datasets beyond its initial training scope.

### 3.3 Segmentation Output

Figure 1 presents a comparative display of tumor segmentation using the SPARC ensemble model on tumor scans within the BraTS GLA, SSA, and PED validation datasets. The figure highlights the precision of the ensemble model in identifying tumor regions, delineating boundaries distinctly aligned with ground truth annotations. When examining the third row of Figure 1, which represents results from the BraTS PED dataset, SPARC efficiently segments the enhancing tumor (ET), depicted in green. However, a classification challenge is noted where Edema is erroneously identified as a non-enhancing tumor (NC), indicated in red. This specific error aligns with common automated segmentation challenges detailed in [10], particularly prevalent in pediatric tumor analysis due to their distinct morphological characteristics.

Such misclassification affects SPARC's performance metrics in the tumor core (TC) category within the PED dataset, as reflected in Table 1. The TC comprises both the ET and NC, and inaccuracies in classification can lead to decreased performance scores, elucidating the need for refined algorithmic precision tailored to pediatric oncological imaging.

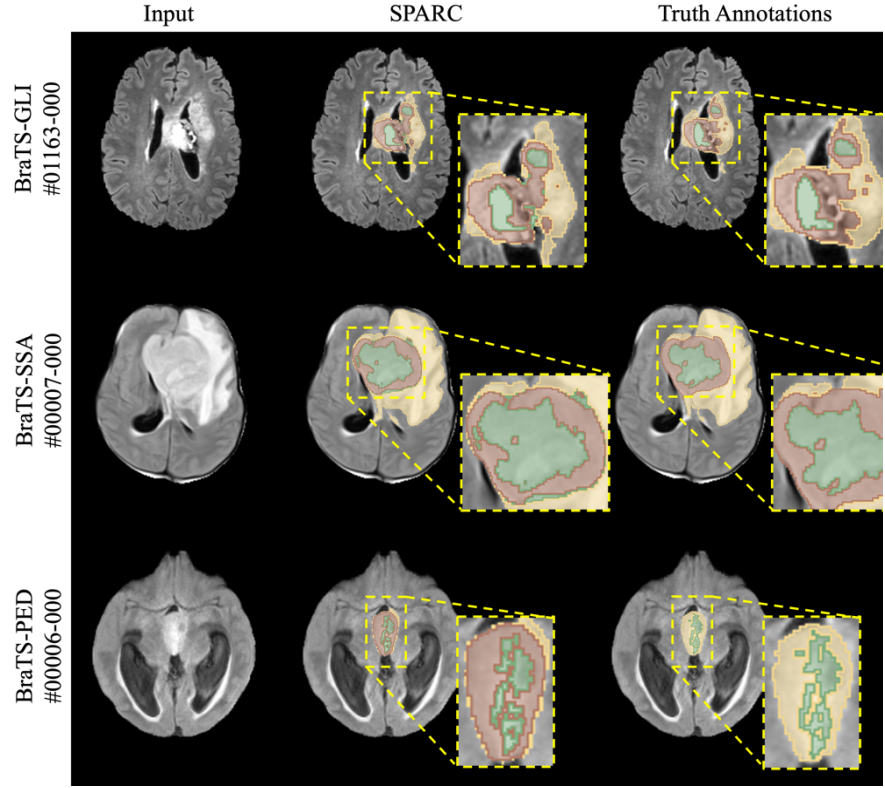

**Fig. 1.** Visual segmentation outputs by our ensemble model for BraTS 2023 GLA (*up*), SSA (*middle*), and PED (*down*) sets. Tumor labels are Edema in *yellow* (ED), Enhancing Tumor in *green* (ET), and Non-enhancing Component (NC) in *red*.

### 3.4 Inpainting Output

Our enhanced 3D Pix2Pix GAN model was evaluated in the BraTS 2023 Inpainting Challenge, where it achieved good results. Table 3 details the performance metrics, including MSE, PSNR, and SSIM. Specifically, the model achieved an MSE of 0.0533, PSNR of 16.4413, and SSIM of 0.6956 for the validation set, while it showed improved performance on the test set, with an MSE of 0.0665, PSNR of 17.2619, and SSIM of 0.7242. These findings affirm the robustness and generalizability of the proposed inpainting model.

Visual representations provided in Figure 2 augment the quantitative data, where various stages of the inpainting process are shown. Each row presents a phase in the inpainting process: the top row features voided T1 scans, the middle row shows the regions masked for inpainting, and the bottom row displays the predictions generated by our model. The enhanced GAN model demonstrates particularly effective inpainting for smaller tumor regions, where the predictions closely align with the expected healthy tissue patterns. The high-resolution generated samples depict a high degree of structural accuracy while showing seamless reconstruction of complex brain tissues, reaffirming the quantitative outcomes reported in Table 3. These results underscore the potential of our approach for clinical applications, providing a valuable tool for medical professionals in the assessment and planning of brain tumor treatments.

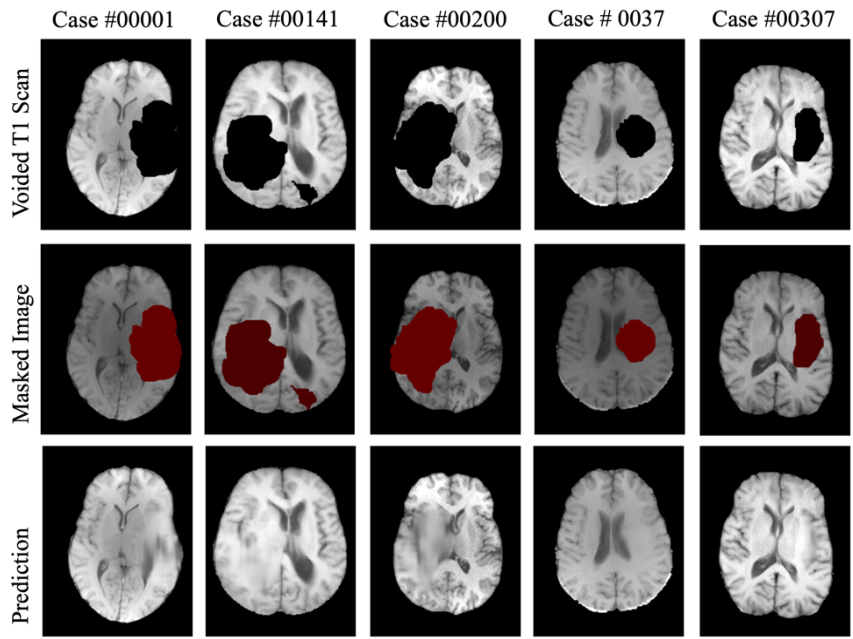

**Fig. 2.** Enhanced Pix2Pix GAN model predictions on 2D transversal T1 MRI slices.

**Table 3.** Results of the enhanced 3D Pix2Pix model on the BraTS 2023 Inpainting Challenge.

| Dataset | MSE | PSNR | SSIM |
| --- | --- | --- | --- |
| Valid | 0.0533 | 16.4413 | 0.6956 |
| Test | 0.0665 | 17.2619 | 0.7242 |

## 4  Discussion

The extensive evaluations of our models within the BraTS 2023 challenges have unveiled the strengths and potential limitations of our proposed methodologies. Through these analyses, we have elucidated the capabilities of our ensemble learning model, SPARC, and the enhanced 3D Pix2Pix GAN, across the diverse datasets of GLA, SSA, and PED.

The results, as presented in Table 1, outline the performance metrics across the different models: TransBTS, nnU-Net, DeepSCAN, and SPARC model. The nnU-Net model exhibits robust performance in terms of lesion-wise DSC scores, outperforming TransBTS and DeepSCAN in most metrics. Notably, our SPARC model, leveraging post-processing strategies, shows promising results, particularly in the enhancing tumor sub-region. This underscores the efficacy of ensemble learning in enhancing the robustness and generalization capabilities of our segmentation models.

Furthermore, the scores of SPARC on the PED and SSA tasks are presented in Table 3. These results demonstrate competitive DSC values, indicating that our model can achieve accurate segmentations across varying clinical and imaging environments. However, as the PED dataset analysis suggests, segmentation challenges remain, particularly in pediatric cases where distinguishing between tumor types can be difficult due to their varied morphology. This is evidenced by the misclassification of ED as NC, an issue that resonates with common errors in automated pediatric brain tumor segmentations, as outlined in the literature [10].

Additionally, this methodology, set within the context of the BraTS inpainting challenge, signifies a notable advancement in the field of medical image analysis. By integrating cutting-edge 3D Pix2Pix GAN with our bespoke DeepSeg model—optimized for the specific nuances of the BraTS dataset—we have established a comprehensive framework for the effective inpainting, and segmentation of glioma-affected brain regions as listed in Table 3. For an in-depth exploration of the techniques and their implications, we direct readers to the detailed publications [13, 21].

Despite the promising results, challenges remain, such as optimizing post-processing strategies to balance the trade-off between false positives and true positives. Additionally, addressing the domain gap between training and test datasets, particularly evident in the BraTS challenge, remains an essential area for improvement. Further exploration of advanced data augmentation techniques and normalization strategies, as well as investigation of more complex architectures and domain adaptation methods, could be beneficial for enhancing our model's performance in the future.

In conclusion, our methodology for the BraTS challenges of 2021 and 2023 represents a comprehensive effort to advance the field of medical imaging for brain tumor analysis. By integrating ensemble learning for segmentation with GAN-based techniques for inpainting, we have outlined a robust framework capable of addressing some of the most pressing challenges in neuro-oncology imaging. Our approach demonstrates the potential of combining multiple AI techniques for medical image analysis and also sets a new benchmark for future research in the domain.